\documentclass[10pt,letterpaper,twocolumn]{article} 
\usepackage{ol2}
\usepackage[draft]{hyperref}
\usepackage{amsmath}

\begin{document}
\twocolumn[ 
\title{An optical isolator using an atomic vapor in the hyperfine Paschen-Back regime}
\author{L Weller,$^{1,*}$ K S Kleinbach,$^{1}$ M A Zentile,$^{1}$ S Knappe,$^{2}$ I G Hughes$^{1}$ and C S Adams$^{1}$}
\address{$^{1}$Joint Quantum Centre (JQC) Durham-Newcastle, Department of Physics, Rochester Building, Durham University, South Road, Durham, DH1~3LE, United Kingdom\\
$^{2}$Time and Frequency Division, the National Institute of Standards and Technology, Boulder, Colorado 80305\\
$^{*}$Corresponding author: lee.weller@durham.ac.uk}
\begin{abstract}
A light, compact optical isolator using an atomic vapor in the hyperfine Paschen-Back regime is presented.  Absolute transmission spectra for experiment and theory through an isotopically pure $^{87}$Rb vapor cell show excellent agreement for fields of 0.6~T.  We show $\pi$/4 rotation for a linearly polarized beam in the vicinity of the D$_{2}$ line and achieve an isolation of 30~dB with a transmission $>$~95~$\%$.       
\end{abstract}
] 
\noindent 
Optical isolators are fundamental components of many laser systems as they prevent unwanted feedback.  Such devices consist of a magneto-optic active medium placed in a magnetic field such that the Faraday effect can be exploited to restrict the transmission of light to one direction.  For an applied axial field $B$ along a medium of length $L$, the Faraday effect induces a rotation $\theta$ for an initially linearly polarized light beam, where $\theta = VBL$, and $V$ is the Verdet constant.  An optical isolator is realized when such a medium is positioned between two polarizers set at $\pi$/4 to each other, with an induced rotation of $\theta=\pi$/4; this arrangement provides high transmission in one direction and isolation in the other.

The technologies of atomic Micro-Electro-Mechanical Systems (MEMS) will eventually be required to create lighter and more compact components~\texttt{\cite{dong2010review}} for use in free-space laser communications, ocean measurements and telecommunications.  Currently there is much interest in small, reliable low power laser systems (VCSEL)~\texttt{\cite{tatum2007vcsel}}, fabrication of chip-sized alkali-vapor cells~\texttt{\cite{liew2004microfabricated}} and gas atoms in hollow core fibers~\texttt{\cite{benabid2005compact,ghosh2005resonant}} for atomic frequency references~\texttt{\cite{knappe2007advances}} and magnetometers~\texttt{\cite{shah2007subpicotesla}}.  Other applications include gyroscopes~\texttt{\cite{donley2010nuclear}}, laser frequency stabilization~\texttt{\cite{knappe2007microfabricated}} and atomic sensors~\texttt{\cite{lee2011small}} for cold-atom devices, accelerometers and gravimeters.  Here we show that using similar technologies one can envisage a light, compact, high-performance permanent-magnet isolator. 

Isolators require large Verdet constants whilst maintaining a small absorption coefficient $\alpha$, hence the figure of merit (FOM) for an isolator is the ratio $V/\alpha$.  Commercial isolators often use terbium gallium garnet (TGG), with yttrium iron garnet (YIG) also used in the IR region.  Table~$\ref{tableconstants}$ shows the Verdet constants and FOM for TGG, YIG and Rb vapor at 780~nm.  Note that although the Verdet constant of YIG is much larger than TGG, the latter is used at 780~nm as the performance of the former is strongly compromised by the poor transparency of YIG below 1100~nm~\texttt{\cite{bai2003magneto}}. Note also that the Verdet constant and absorption coefficient of the atomic vapor are strongly frequency dependent, unlike the crystal media.  The FOM for Rb vapour is less than TGG, the much higher Verdet constant however allows a more compact design, as the same rotation is achieved over a much shorter optical path.  The frequency dependence of the dichroic and birefringent properties of atomic vapors have also been exploited to realize narrowband atomic filters; see e.g.~\texttt{\cite{beduini2011ultra}} and dichroic beam splitter~\texttt{\cite{abel2009faraday}}.
\begin{table}[t]
\centering
\caption{Verdet constants and FOMs for the three magneto-optic materials: TGG, YIG and Rb vapor (this work), at a wavelength of 780~nm.}
\begin{tabular}{lccc}
\hline
Material                                  & $V$ (rad~T$^{-1}$~m$^{-1}$)& FOM (rad~T$^{-1}$)     \\\hline	 	
TGG~\texttt{\cite{villora2011faraday}}    & 82		                     & 1~$\times$~10$^{3}$    \\ 
YIG~\texttt{\cite{bai2003magneto}}        & 3.8~$\times$~10$^{2}$ 		 & 2.5      	            \\
Rb vapor                                  & 1.4~$\times$~10$^{3}$      & 1~$\times$~10$^{2}$    \\
\hline
\label{tableconstants}
\end{tabular}
\end{table} 
     
The absolute susceptibility of an atomic vapor can be accurately modeled allowing one to predict both the absorption~\texttt{\cite{siddons2008absolute,kemp2011analytical}} and rotation~\texttt{\cite{siddons2009off,siddons2009gigahertz}}.  Furthermore the model can be extended to include binary-collisions~\texttt{\cite{weller2011absolute}}, and an axial magnetic field~\texttt{\cite{weller2012Stokes}}.  In this Letter we use the model for absolute susceptibility for the Rb D$_{2}$ line to predict the performance of an optical isolator based on a compact magnet producing a field of the order of 0.6~T.  At such fields the vapor becomes transparent to light resonant with the D$_{2}$ line due to large Zeeman shifts.  We measure rotation and transmission and confirm the theoretical predictions and thereby demonstrate the feasibility of using resonant atomic media for optical isolation.  Although here we focus on the Rb D$_{2}$ line, using the same principle, resonant atomic isolators can be implemented for other atomic vapors.  

\begin{figure}[t]
\centering
\includegraphics*[width=0.47\textwidth]{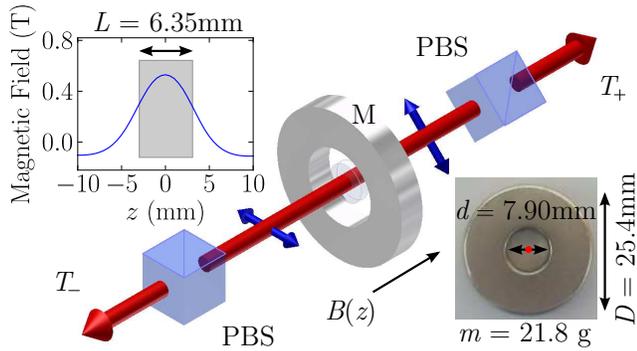}
\caption{Schematic of the experimental apparatus and measured magnetic field profile through the center of the neodymium magnet.  The dimensions and mass of a typical magnet used in this investigation are also shown.  A beam passes through a polarization beam splitter (PBS) providing linearly polarized light along the horizontal axis.  The beam then passes through a heated micro-fabricated cell held in a magnet (M) which provides an axial field.  A second PBS is set to $\pi$/4 to allow high transmission, $T_{+}$, along ${+z}$ and low transmission, $T_{-}$, along ${-z}$.}
\label{MainSetup}
\end{figure}

Figure~\ref{MainSetup} shows a schematic of the experimental apparatus along with details of the neodymium magnet.  An external cavity diode laser was used to scan across the Rb D$_{2}$ transition (5$^{2}$S$_{1/2}$~$\rightarrow$~5$^{2}$P$_{3/2}$) at a wavelength of 780~nm.  After passing through a polarization beam splitter (PBS) the output beam was linearly polarized along the horizontal direction with a 1/e$^{2}$ radius of 80~$\mu$m.  The method adopted for calibrating the frequency axis is described in~\texttt{\cite{weller2012Stokes}}.  A weak probe beam~\texttt{\cite{siddons2008absolute,sherlock2009weak}} traverses a 1~mm heated cell containing isotopically pure $^{87}$Rb and buffer gas with a pressure of several Torr~\texttt{\cite{weller2012paschenback}}.  An aluminium holder with the same design as in~\texttt{\cite{weller2012Stokes}} was used to hold a neodymium magnet, where the cell was held in an oven allowing the laser beam to pass through.  The field profile of the neodymium magnet was measured with a Hall probe and is shown in figure~$\ref{MainSetup}$.  Over the length of the cell the field was uniform at the 2~$\%$ level.  After traversing the cell and magnet, a second PBS was set to $\pi$/4  allowing high transmission along $+z$ for an induced rotation of $\pi/4$.  With this arrangement one would also expect isolation along $-z$. 

\begin{figure}[t]
\centering
\includegraphics*[width=0.47\textwidth]{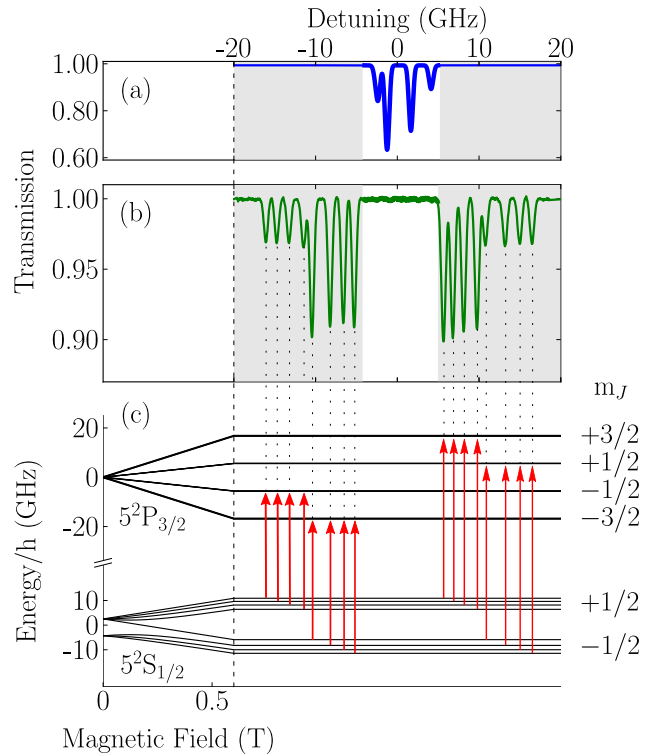}
\caption{Transmission spectra of atomic vapour (a) without and (b) with an applied field illustrating the opening of a transparency window over resonance.  Plot~(a) shows the theoretical (solid blue) transmission spectra at a temperature of 60.4~$^{\circ}$C through a natural-abundant cell, highlighting the four absorption peaks of interest for isolation.  Plot~(b) shows the measured (solid green) line transmission spectra at a magnetic field of 0.576~T and a temperature of (60.4 $\pm$ 0.2)~$^{\circ}$C through a $^{87}$Rb cell.  Plot~(c) shows the energy level splittings for the ground (5$^{2}$S$_{1/2}$) and excited terms (5$^{2}$P$_{3/2}$) of $^{87}$Rb on the D$_{2}$ line.  In the hyperfine Paschen-Back regime m$_{J}$ is now a good quantum number, the transitions with $\Delta$m$_{J}$ = $\pm$1 correspond to the 16 absorption peaks we measure in plot~(b).}
\label{energydiagram}
\end{figure}

Figure~\ref{energydiagram} shows the absolute transmission spectra for the Rb D$_{2}$ line through (a) a natural-abundant (72$\%$ $^{85}$Rb, 28$\%$ $^{87}$Rb) cell in the absence of an applied magnetic field and (b) a $^{87}$Rb cell in the presence of an applied magnetic field of 0.576~T.  For this field (the hyperfine Paschen-Back regime~\texttt{\cite{sargsyan2012hyperfine}}) the Zeeman shift is large compared with the hyperfine splitting of the ground and excited terms, and m$_{J}$ is the good quantum number.  In (b) we observe the 16 absorption peaks corresponding to the $\Delta$m$_{J}$ = $\pm$1 transitions.  Magnetic fields of such magnitude force a large splitting in the transition frequencies, giving a region of high transmission and large dispersion where we would normally expect absorption on the Rb D$_{2}$ line: this is the basis for our isolator. 

\begin{figure}[t]
\centering
\includegraphics*[width=0.47\textwidth]{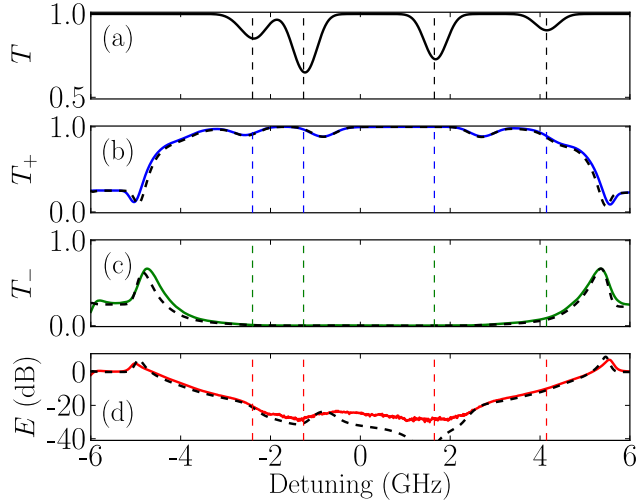}
\caption{Forward $T_{+}$ and backward $T_{-}$ transmission illustrating the isolator effect as a function of detuning around resonance.  Plot~(a) shows the theoretical (solid black) transmission through a natural-abundant cell in the absence of field and at a fixed temperature of 60.4~$^{\circ}$C.  The four absorption peaks highlight the required detuning values for isolation.  Plots~(b), (c) and (d) show comparison between experiment (solid colored) and theory (dashed black) through an isotopically pure $^{87}$Rb cell in the presence of a fixed field of 0.597~T and a temperature value of (135.2 $\pm$ 0.4)~$^{\circ}$C, which corresponds to a region where we expect $\pi$/4 rotation.}
\label{Isolation}
\end{figure}

Figure~\ref{Isolation} shows absolute transmission spectra for the Rb D$_{2}$ line.  Plot~(a) shows the theoretical (solid black) transmission through a natural-abundant cell in the absence of field and at a temperature of 60.4~$^{\circ}$C.  Plots~(b), (c) and (d) show comparison between experiment (solid colored) and theory (dashed black) for the rotation of light describing the high transmission along $+z$ (blue), low transmission along $-z$ (green) and extinction values of the isolator (red), respectively.  All three spectra are obtained for an isotopically pure $^{87}$Rb cell with a fixed field of 0.597~T and a temperature value of (135.2 $\pm$ 0.4)~$^{\circ}$C, which corresponds to a region where we expect $\pi$/4 rotation.  These spectra demonstrate the narrow-band nature of isolators based on atomic vapors as there is only a constant dB over a range of 4~GHz.  The increased structure in the $T_{+}$ signal which is present in the theoretical and measured signals is the result of additional weak absorption features owing to the fact that the ground terms are not completely decoupled; a detailed study of these features will be the topic of a future publication~\texttt{\cite{weller2012paschenback}}. 
    
Important characteristics for isolators are their: ability to extinguish backscattered light, power threshold and temperature stability.  In figure~\ref{Isolation} we define the isolation of the device as $E = -10 \log(T_{+}/T_{-})$~dB, where $T_{+}$ is the transmitted light along ${+z}$ and $T_{-}$ is the transmission along ${-z}$.  Previous examples of crystal isolators have measured extinctions of 47~dB for a single device~\texttt{\cite{gauthier1986simple}} and 60~dB for a back-to-back device~\texttt{\cite{wynands1992compact}}.  To the best of our knowledge this is the first resonant atomic vapor isolator and we achieve a 30~dB suppression, limited by the extinction of our polarizers.  The excellent agreement between theory and our model is typically achieved in the weak-probe regime.  However, when the power of the probe beam was increased by 6 orders of magnitude the extinction in figure~\ref{Isolation} changed by less than 8~dB.  Owing to the strong temperature dependence of the birefringent properties of the medium a device utilizing this effect would require the vapor cell to be temperature stabilized within $\pm$~0.2~$^{\circ}$C.

In summary, we have demonstrated the principle of an optical isolator for the Rb D$_{2}$ line by exploiting the magneto-optical properties of an isotopically pure $^{87}$Rb vapor in the hyperfine Paschen-Back regime.  We show $\pi$/4 rotation for a linearly polarized light in the vicinity of the D$_{2}$ line and achieve an isolation of 30~dB.  
This work is supported by EPSRC.  We thank James Keaveney for the design of the cell heater.  

\pagebreak
\section*{Informational Fourth Page}

\end{document}